# Understanding the loss in community resilience due to hurricanes using Facebook Data

**Tasnuba Binte Jamal** (corresponding author)
Department of Civil, Environmental, and Construction Engineering
University of Central Florida, Orlando, Florida 32816, USA
Email: TasnubaBinte.Jamal@ucf.edu

**Samiul Hasan, Ph.D.**
Department of Civil, Environmental, and Construction Engineering
University of Central Florida, Orlando, Florida 32816, USA
Email: Samiul.Hasan@ucf.edu

## ABSTRACT

Significant negative impacts are observed in productivity, economy, and social wellbeing because of the reduced human activity due to extreme events. Community resilience is an important and widely used concept to understand the impacts of an extreme event to population activity. Resilience is generally defined as the ability of a system to manage shocks and return to a steady state in response to an extreme event. In this study, aggregate location data from Facebook in response to Hurricane Ida are analyzed. Using changes in the number of Facebook users before, during, and after the disaster, community resilience is quantified as a function of the magnitude of impact and the time to recover from the extreme situation. Based on the resilience function, the transient loss of resilience in population activity is measured for the affected communities in Louisiana. The loss in resilience of the affected communities are explained by three types of factors, including disruption in physical infrastructures, disaster conditions due to hurricanes, and a community's socio-economic characteristics. A greater loss in community resilience is associated with factors such as disruptions in power and transportation services and disaster conditions. Socioeconomic disparities in loss of resilience are found with respect to a community's median income. Understanding community resilience using decreased population activity levels due to a disaster and the factors associated with losses in resilience will enable us improve hazard preparedness, enhance disaster management practices, and create better recovery policies towards strengthening infrastructure and community resilience.

## 1. INTRODUCTION

Many countries in the world are now facing major disasters such as wildfire, tornado, hurricane, tropical storm, and flooding. In the USA, total cost of the damages due to weather and climate disasters exceeded $2.295 trillion since 1980 (*1*). From 2000 to 2021, there are a total of 28 major hurricanes in the USA (*2*) and the induced damages have significantly increased due to major landfalls in recent years (*3*). For instance, Hurricane Irma caused a damage of about $50 billion in Florida (*4*). Damages by such extreme events cause a change in regular human activities. Compared to regular periods, human activities in disaster periods go through a significant amount of perturbation. People are less likely to work or move the same way in disaster situations as they do in normal conditions. Since a decrease in human activity is an indication of drop in business, recreation, and health services (*5*) (**Fig. 1**), understanding the changes in human activities is a key factor to analyze the hardships to the daily life of disaster affected communities.

In general, resilience indicates the ability of a system to return to its normal state or situation after a disruption due to an extreme event (*6*, *7*). At the onset of a hurricane, human activities start to decrease in the affected area, reach maximum drop after a certain time, and then start to recover. Since human activities are not at usual level, the amount of decrease in activities compared to regular periods can be termed as the loss of resilience for the disaster affected community. To quantify community resilience in human activity, different types of location data (Twitter and mobile phone location data) have been used previously (*7*, *8*). Technological advancements provide researchers the access to high-fidelity location data. Such high-resolution location data including taxi data (*9*, *10*), GPS data (*11*, *12*), cell phone call recordings (*13*), Wi-Fi (*14*), and mobile phone datasets (*15*, *16*) were used previously to understand human mobility and activity following an extreme event. However, these datasets are not always available; they are often proprietary, and sometimes confined to some specific point of interests (POIs) only (e.g., shopping mall, restaurants, etc.). To effectively quantify community resilience, data must be easily accessible and usable (*17*). These data unavailability issues can be avoided if location data from a widely used service are made accessible. Some studies have utilized post-disaster survey data to understand population activity due to natural disasters (*18*–*23*). However, such survey data are not reliable because they cannot capture the dynamic patterns of recovery and respondents may not remember everything. So, it is challenging to collect longitudinal data from disaster-affected regions through post-disaster surveys (*15*, *24*).

To this end, Facebook Data for Good (now 'Data for Good at Meta') platform is sharing aggregate data on where people are located before, during, and after a crisis event following a crisis event to help humanitarian organizations (*25*). Data from 'Facebook Data for Good' platform are free and easily available. This platform shares the counts of Facebook users who enable location services on their mobile device (*25*). Whenever there is a disaster, such data of Facebook users are globally available from the affected regions. Facebook population data is a great source to identify crisis events over a certain time window and investigate the impacts of these events to population activity. Worldwide, 2.9 billion monthly active Facebook users are reported in 2022. About 240 million users were found in the USA in 2021 (*26*). According to Pew Research Center's national survey (2019), Facebook usage rates are high in the USA (*27*). For example, 69% of U.S. adults use Facebook as a social media platform (*27*), indicating the applicability and reliability of Facebook data in research.

This study investigates community resilience in population activity against a hurricane utilizing Facebook data before, during and after Hurricane Ida. We define community resilience as the ability of an affected community to return to its regular activity (in terms of social activity). So, the loss of resilience is equivalent to the amount of disruption in population activity. A resilient community can withstand an extreme situation while a non-resilient or vulnerable community undergoes a significant and prolonged disruption. Therefore, a resilient community loses a small amount of resilience and a vulnerable community on the other hand loses a significant amount of resilience due to disasters. In general, resilience indicates a community's long-term property in reaction to all potential crisis events. Resilience loss is referred in this paper as the "*transient loss of resilience*" because it is determined in reaction to a single disaster (*7*).

This paper is a first step towards developing methodologies to determine the loss in community resilience from Facebook data with the following specific contributions:

1. This study demonstrates the use of a large-scale macroscopic location dataset collected from Facebook to quantify community resilience. While data from this source have been used in the field of ubiquitous computing and research on migration and evacuation, we add a new dimension to this type of data by quantifying resilience for an affected community due to an extreme event such as a hurricane.

2. It further develops a statistical model to investigate the association between multiple types of infrastructure disruptions (transportation and electricity services) and the transient loss of resilience in population activity due to hurricanes, while accounting for hazard characteristics. While previous studies investigated disparities in community resilience over varying socioeconomic and demographic attributes, we add a new perspective by studying the impacts of multiple types of infrastructure disruptions and disaster conditions on community resilience.

When a disaster occurs, the socio-infrastructure systems of a community might be significantly disrupted such as disruptions in electricity services, transportation networks' functionality, business activities, delayed and inequal assistances from Govt. and non-Govt. organizations. Such disruptions preclude population activity from coming back to normal state and prevent communities from being resilient to disasters (**Fig. 2**). It is important to investigate which factors are associated with the transient loss of resilience of an affected community. **Fig. 2** shows a conceptual framework of the factors that are likely to be associated with a community's loss of resilience. This study assesses the combined effects of physical infrastructure damage, disaster condition, and socio-economic characteristics on loss of resilience at county-subdivision level (see **Fig. 2**). Understanding the drop in population activity levels due to a disaster, its gradual recovery processes, and the factors associated to such processes will allow us to improve hazard preparedness, enhance disaster management practices, reduce economic losses, and create better recovery policies. The findings of this study provide insight into effective identification of less resilient communities to hurricanes from large-scale, real-time, free, and easily accessible Facebook data, as well as the correlates of transient loss of community resilience. This study shows how disaster condition and disruption in multiple physical infrastructures are associated with the transient loss of community resilience and highlights that disparities in recovery patterns are associated with socioeconomic attributes.

## 2. LITERATURE REVIEW

Previous research analyzed location-based data to understand recovery patterns and quantify the losses to the lifestyle of an affected community due to disasters. Analyzing large-scale location-based datasets (cell phone call recordings and social media posts), studies found that recovery patterns are not random, they follow some specific patterns (*10*, *11*, *16*, *28*, *29*). Different types of high-resolution location datasets were used previously to understand human mobility and activity following an extreme event (*9–16*). We can now measure the recovery trajectories using big data at previously unheard-of high frequency, granularity, and scale. Such big data enables us to further quantify the fundamental resilience features of communities utilizing data driven complex systems modeling (*30*). For a more effective, inclusive, and responsive disaster response and recovery; mobile phone location data holds enormous potential (*31*). Through the use of mobile phones, it is now possible to collect spatio-temporally detailed observations of individual mobility throughout a vast region (*32*, *33*) and it was discovered that human trajectories exhibit a high degree of temporal and spatial regularity. Guan et al. (*34*) developed methods to track changes in social interaction using Twitter data and in two transportation networks (subway and taxi) using subway ridership and taxi data on daily basis due to a major disaster. Juhasz et al. (*35*) investigated the effect of Hurricane Irma on visitation numbers in Florida considering six different point of interest (POI) provided by SafeGraph platform. They identified factors associated with increased or decreased distance between home and a specific POI category. Sudo et al. (*12*) proposed a particle filter method to predict human mobility several hours ahead of an event using real-time location data. Yabe et al. (*16*) focused on population recovery patterns during post-disaster periods, by observing human mobility trajectories of mobile phone users. They explained the heterogeneity in displacement rates and the speed of recovery across communities at local government units (LGU) level (LGUs correspond to counties in the USA).

Overall, previous studies focused on recovery trajectories, human mobility and activity patterns, displacement rate/systemic impact (e.g., percentage of population impacted by a disaster) and recovery time of the affected community from different types of location datasets. Despite such progress, the current body of literature needs a general term for understanding population displacement rate and recovery patterns after disasters with easily accessible, free, and representative longitudinal data. For example, displacement rate or recovery speed separately may not reveal the extent of disruptions to the activities of a community by a disaster since despite having small rate of displacement, it may take longer time for them to recover or vice-versa. To better understand the impacts of a disaster to an affected community, we need to consider 'resilience' which involves both systemic impact and duration of impacts (*7*, *8*, *15*, *36*, *37*). In general, resilience indicates the ability of a system to return to its normal state after a disruption (*6*, *7*). For community resilience, it is usually defined as the ability of a disaster affected community to come back to the normal life. Hong et al. (*8*) and Roy et al. (*7*) quantified community resilience in population activity and mobility using geo-located mobile device and social media data, respectively. However, these studies are limited to only quantifying community resilience. They did not investigate the factors associated with the loss of resilience of a community and how recovery patterns vary across affected communities despite facing similar levels of shocks.

Several studies investigated the disparities in disaster response and recovery patterns associated with varying socioeconomic, demographic, and geophysical attributes (*8*, *38*, *39*). However, few studies focused on the effects of multiple infrastructure disruptions and disaster conditions on community resilience in population activity. Yabe et al. (*16*) considered the effects of median income, population size, connectedness to cities, and durations of power outage on recovery speed and displacement rates after hurricanes. Sadri et al. (*40*) focused on the effects of physical infrastructure damage, social capital, household characteristics, and recovery assistance on recovery time of a household due to tornados using traditional survey data in southern Indiana. The importance of transportation networks' recovery and disaster conditions of the regions were understudied in the current literature, despite they have significant implications on policymaking for disaster affected communities. Podesta et al. (*15*) showed the relationship between inundation with hazard impact and the restoration time for community to get back to their regular activity in Houston, Texas due to Hurricane Harvey. Again, these studies focused either on recovery time or percentage of population impacted due to disaster instead of considering a general single term (e.g., the loss in community resilience). Moreover, the impact of major infrastructure systems (e.g., transportation network) and the severity of hazard (e.g., wind speed) on community resilience were not investigated in these studies. After an extreme event, infrastructure systems are critical to recover to maintain the well-being of a community (*41*). As such, how infrastructure disruptions are associated with community resilience needs to be understood to enhance recovery policies. For instance, Yabe et al. (*42*) found that expanded centralized infrastructure systems of cities can enhance the recovery efficiency of critical services from the study on Hurricane Maria. Resilience increases when facilities are distributed more fairly (*43*), and soft and local policy toolkits are adopted (*44*).

Besides understanding recovery trajectories after disasters, location datasets were used to study migration, evacuation and disease spreading. Mobile phone datasets were used for inferring migration patterns (*45–47*), and disease spread (*48–50*). Similarly, Facebook data was used to study migration (*51*) and evacuation (*52*). Acosta et al. (*51*) estimated population changes due to migration over the course of a year after Hurricane Maria in Puerto Rico using Facebook data and found a 17% decrease in population in 2017. Fraser (*52*) used trajectories of Facebook users' movement to analyze evacuation pattern at county subdivision level due to Hurricane Dorian in Florida. He found that linking social capital and soft community-focused preparation strategies increased evacuation across cities. By comparing evacuation patterns from 10 different hazards in the US and Japan from 2019 to 2020, Fraser (*53*) showed that some disasters have more similar evacuation patterns than others. After analyzing mobile phone users' positions from four major earthquakes, Yabe et al. (*54*) discovered that an individual's evacuation likelihood is strongly associated with the seismic intensity they experience.

Previous literature suggests several challenges in quantifying community resilience. First, it is challenging to collect longitudinal data from disaster-affected regions through post-disaster surveys. Although location data can be used as an alternative to surveys, most of the location datasets are proprietary (e.g., mobile phone data) or have small sample sizes (e.g., social media posts). To effectively quantify community resilience, data must be easily accessible and usable (*17*) by decision makers and emergency officials. These data-related challenges limit a wider adoption of previous approaches quantifying resilience. Second, previous studies mainly focused on resilience quantification without answering why some communities were less resilient compared to others. Some studies focused on socio-economic inequality in hurricane impact analysis. Besides socio-economic perspectives, other factors such as hurricane characteristics and infrastructure disruptions can affect population activity. For example, the importance of transportation networks in the recovery from a disaster were understudied in the literature. Third, previous studies considered either recovery time or maximum drop in population activity as a dimension of interest in the developed statistical models. However, measuring the loss of resilience gives more information of hurricane impact since it considers both recovery time and maximum drop (*7*, *8*, *15*, *36*, *37*). More specifically, the following research questions are yet to be answered:

**RQ 1**. *How to quantify the loss of community resilience from an easily accessible, free, and representative longitudinal dataset?*

**RQ 2.** *How can the heterogeneity in loss of community resilience be explained using different types of infrastructure disruptions, disaster condition while accounting for socioeconomic attributes?*

To answer these questions, this study shows how aggregate location data from Facebook can be used to measure and understand community resilience after a disaster. We apply the concept of resilience for understanding population activity under Hurricane Ida at a county-subdivision level. This study develops a statistical model to explain the heterogeneity in loss of community resilience. Since this study analyzes the data at a finer geographical level (in Louisiana 64 parishes are divided into 579 county subdivisions), the findings ensure better understanding of community resilience and the association between loss of community resilience and physical infrastructure disruptions, disaster condition and socio-economic aspects. The findings from this study can aid policy makers and emergency officers to identify and strengthen less resilient communities to hurricanes and thus to support their disaster preparedness activities.

## 3. DATA DESCRIPTION
### 3.1 Facebook Population Data

For this study, Facebook population data was used at an administrative region level collected from Facebook's Data for Good platform ([dataforgood.facebook.com/dfg/about](dataforgood.facebook.com/dfg/about)) (currently known as Data for Good at Meta). Facebook Population data shares the aggregate number of Facebook mobile app users who enable location services in their mobile devices (*25*). It is to be noted that this dataset does not depend on Facebook usage by the users. That is, even if people do not use Facebook after coming back to the original home location after the hurricane, as they will be distressed and burdened with all kinds of recovery activities, Facebook can still record users' location. This dataset provides the average number of users present in a region during the baseline period (90 days before the day the data was generated), the number of users during a crisis event, and the difference between these two quantities. Additionally, a *z* score is provided to highlight the areas with the most significant differences between regular and crisis periods. The *z* score is calculated by [(users during crisis – mean baseline users)/ variance of baseline users] with values ranging between -4 and 4.

For our considered time window (from 25$^{th}$ August 2021 to 30$^{th}$ September 2021), data was collected at administrative region level 4 which is equivalent to county-subdivisions. This platform provides data at 8-hour intervals (00:00 UTC, 08:00 UTC and 16:00 UTC) (*25*). We considered only 16:00 UTC to 00:00 UTC; 16:00 UTC indicates 11 a.m. local time in Louisiana. The period from 11 a.m. to 7 p.m. local time corresponds to the peak activity hours for many people. This includes work hours, lunch breaks, and post-work leisure time. Analyzing data during this period can provide insights on how the hurricane impacted population activity in one the most active periods of the day.

Facebook population data provides only the names of the county-subdivisions (if administrative region level is 4) without the names of the counties associated with each county-subdivision. For Louisiana, it is not possible to identify a unique county-subdivision because several county subdivisions under different parishes (equivalent to a county in other states) have the same names. For example, East Baton Rouge, Acadia, Jefferson all these counties have county-subdivision named District 1. To identify the county that a county-subdivision belongs to, Facebook user movement data at an administrative region level 4 was collected from the same platform. This dataset has the latitude and longitude of the center of the boundary polygon shape (e.g., county subdivisions) which is not available in the Facebook population datasets. Using the censusgeocode package in Python, the corresponding county of a county-subdivision was determined from the latitude and longitude of the center of a county-subdivision. Then, the county names were merged with the Facebook population datasets based on polygon ID and obtained the unique county subdivisions. A polygon ID is a unique identifier for a county subdivision provided in these datasets (since the polygon ID is same for a county subdivision in both datasets).

The collected dataset for Hurricane Ida had data for 442 county subdivisions from 73 different counties across Alabama, Mississippi, and Louisiana states. This study focused only on Louisiana state because decreased population activity was observed mostly in Louisiana. People might have evacuated to Alabama and Mississippi and increased activity was observed. Since this study focused on decreased activity, which causes loss of community resilience; data from Alabama and Mississippi was not considered. The final dataset included observations for 327 county subdivisions from 39 parishes in Louisiana where a parish is equivalent to a county of other states in the USA.

**Representativeness of Facebook population data**

In this study, Facebook users were used as a sample of the population. This requires validating whether the Facebook data represents the actual population. Pew Research Center's national survey in 2019 reported high Facebook usage rates in the USA. For example, 69% of U.S. adults use Facebook as a platform or messenger app and 74% of the Facebook users visit it once a day. Facebook usage is also high among men (63%) and women (75%), among White, Black, and Hispanic population (each 69–70%) (*27*).

To further validate this, the correlations between the number of Facebook users and population were calculated; a similar approach was adopted for macroscopic/aggregate location data in previous study (*16*). Pearson's correlations (**Fig. 3**) between the number of Facebook users and population for 39 parishes and for 327 county (parish) subdivisions of Louisiana were found as 0.98 and 0.96, respectively. **Fig. 3** also indicates that the number of Facebook users is linearly proportional to the population both at county subdivision level and parish level. As such, Facebook users can be used as a sample of total population in this study. Facebook population dataset was found to have an average penetration rate of 7.5% to actual population in Louisiana.

**3.2 Physical Infrastructure Data**

This study considered electricity services, transportation network, housing damage and age from physical infrastructure disruption of **Fig.2**.

**Disruption on roads**

Road disruptions data was collected from Regional Integrated Transportation Information System (RITIS) and included in hours for each county subdivision within analysis period (from 25$^{th}$ August to 30$^{th}$ September 2021). RITIS provides event data from three different agencies with each agency having their own definitions of categorizing different events. This study mainly considered weather hazard, weather closures, road closed, closures and obstruction agency-specific event types. For a disruptive event on roads, the dataset had only the latitude and longitude information, which were used to identify the corresponding county subdivision of the event.

**Power service restoration time**

To investigate how disruptions in power services are impacting community resilience, this study included restoration time from power outage of the parishes. Power outage data was collected from Bluefire Studios LLC. for Hurricane Ida. This dataset reported the total number of electricity customers in a parish of Louisiana and the number of customers having power outages during Hurricane Ida in 1-hour intervals from 20$^{th}$ August to 30$^{th}$ September 2021. We used the duration between the time when 10% of customers or more of a particular county first lost their electricity services and the time when 10% of customers or less were yet to restore their power services (**Fig. 4**). It was observed that the counties where less than 10% of customers lost power services, did not take long time to get their electricity services back. Due to data unavailability, it was assumed that restoration time of power outages for all the county subdivisions under a parish is same. On average, it took same time for all the county subdivisions under a parish to restore the power services. Previous studies found significant Moran's I value (*55*), Lagrange Multiplier (LM) and the Robust Lagrange Multiplier (RLM) test statistics (*56*) for percentage of customers without power and restoration time, respectively due to hurricanes, indicating that power outage in neighborhood areas have similarity. Significant values of these test statistics indicate the presence of spatial correlation (clustering) in power outage restoration. This means that areas with longer restoration time are close to each other, and similarly areas with shorter restoration time are close to each other. Previous studies (*57, 58*) also considered restoration strategies from power outages due to a hurricane at county-levels. Thus, it can be reasonably assumed that the variation in restoration times of the county subdivisions of a particular county should be small.

**Property damage data**

To explore if property damages have an impact on community resilience, this study included property damage in terms of average inspected damage (based on Federal Emergency Management (FEMA)'s inspection guidelines). It is available for valid registrations from households within the state, county, zip that had a complete inspection, collected from FEMA's housing assistance datasets for Hurricane Ida. The corresponding county subdivision of an observation was based on city, zip code, parish, and state. For some county subdivisions no damage data was found in the FEMA dataset; we assumed that there was no FEMA inspected damages in those subdivisions.

**Age of the houses**

Since old houses are prone to be damaged by natural disasters, county subdivisions with a greater number of older houses might be less resilient to hurricane. To indicate this variable, the percentages of houses that are built before 2000 (more than 21 years old when Hurricane Ida occurred in 2021) in each county subdivision were collected from American Community Survey (ACS).

**3.3 Disaster Condition**

**Distance to hurricane path**

Hurricane Ida's path was collected from National Hurricane Center (NHC). Haversine formula was used (**Equation 1**) to calculate the distance between the center of a county subdivision and hurricane path and the minimum distance from the center of county subdivision to hurricane path was considered. This formula is used to calculate geographic distance on earth between two different latitude – longitude values of two different points on earth, giving the shortest distance between two points on earth surface (*7*). We considered hurricane path from 26$^{th}$ August to 4$^{th}$ September because hurricane was dissipated after 4$^{th}$ September 2021.

$$d = 2r \, arcsin \left( \sqrt{sin^2 \left( \frac{\phi_2 - \phi_1}{2} \right) + cos \, \phi_1 \, cos \, \phi_2 \, sin^2 \left( \frac{\lambda_2 - \lambda_1}{2} \right)} \right) \qquad (1)$$

where, $\phi_2$, $\phi_1$ are the latitude of point 1 and latitude of point 2, $\lambda_1$, $\lambda_2$ are the longitude of point 1 and longitude of point 2, and r is the radius of earth.

As disaster conditions, other variables such as the type of evacuation orders issued (mandatory and voluntary) and flood depth (**Fig. 2**) could be used. However, these variables are likely to be correlated with the distance from hurricane path because the regions close to the hurricane path are likely to issue a mandatory or voluntary evacuation order for their residents. Besides, these variables are difficult to obtain at a county subdivision level.

### 3.4 Socio-economic characteristics

As socio-economic characteristics, this study included the median household income, the percentage of Black population, and the percentage of Hispanic population in each county subdivision. Percentage of Hispanic population and poverty information for the county subdivisions were collected from the demographic and economic characteristics of American Community Survey (ACS) 5-Year Data Profile for 2020. About 33% of the populations in Louisiana were Black which was the 2nd highest population group and about 5.6% of the populations were Hispanic which was the 3rd highest population group (*59*).

The list of candidate variables for the model and their descriptive statistics are provided in **Table 1**. The correlations between these candidates were tested using Pearson correlation coefficient measure (**Fig. 5**). No highly correlated variables were identified, but moderate correlation between restoration time of power outage and distance to hurricane path was found (*60*). For the housing age variable, the variance inflation factor (VIF) was 11. For the remaining variables, VIF was less than 7. The multicollinearity condition number was 3.013 (which was below 30), indicating that collinearity should not be an issue with statistical models.

## 4. METHODS
### 4.1 Quantifying Community Resilience

To quantify community resilience, transient loss of resilience was calculated using **Equation 2**. Bruneau et al. (*36*) proposed this equation in the context of infrastructure resilience due to an earthquake and it was later adopted by Roy et al. (*7*) for quantifying transient resilience loss (TRL) in human mobility.

$$TRL = \int_{t_0}^{t_1}[1 - Q(t)]dt \tag{2}$$

where TRL denotes transient resilience loss, $Q(t)$ denotes a quality function of a system at time $t$, and $(t_1 - t_0)$ is the recovery time. **Fig. 6** shows a conceptual diagram to illustrate these terms. The area between the horizontal dashed line (baseline value) and decreased quality function (solid line) from $t_0$ to $t_1$ is defined as transient loss of resilience of a system. The horizontal dashed line indicates that the performance of a system is supposed to follow this line if the system does not experience any disruption. The solid curved line within $t_0$ to $t_1$ indicates system performance follows this trend due to the occurrence of an extreme event. So, the resilience is the area under the quality function curve from time $t_0$ to $t_1$. It can be obtained by subtracting the transient loss of resilience from the area under horizontal line from $t_0$ to $t_1$.

This study used Facebook population data to quantify community resilience against hurricanes. First, for each community, population activity rate available from Facebook data was assumed as a measure of its quality function $Q(t)$. For a given day, population activity rate was defined as the rate of Facebook users who were found in a given county-subdivision out of all the affected users on that day. The population activity rate on a given day was calculated by dividing the number of users observed in a particular county-subdivision on that day by the total number of typical users represented by the number of average Facebook users observed 90 days prior to that day (i.e., baseline population).

We calculated Transient resilience loss (TRL) (**Equation 2**) by a numerical integration method. The analysis period was from 25th August 2021 to 30th September 2021 (37 days). This time window was selected because most of the activity fluctuation curves for the affected county subdivisions returned to the normal state within this time. Since this study was concerned with loss of community resilience, any increase in activities from the base period (if there was any within this time window) was not considered. Among the given 327 county subdivisions in this data source, we did not consider the county subdivisions that did not have any drop in population fluctuations curves. We did not consider the county subdivisions for which population activity rate did not fall below the 90% (100% means population activity did not decrease at all). In other words, we considered the county subdivisions where 10% or more Facebook users were missing at least for one day within 29th August to 5th September compared to the baseline Facebook users. Hurricane Ida had its landfall on 29th August 2021, and we considered county subdivisions where significant drop in Facebook users was observed within 7 days of landfall. In this way, observations for 166 county subdivisions out of 327 were obtained. To validate that there was extreme situation in the selected 166 county subdivisions, we further checked the Z- scores provided in Facebook population datasets and for all of them, the minimum Z-score was less than -1.82 (>45% of -4) for more than 1 day out of our considered 37 days.

To obtain resilience of an affected community, we subtracted the transient loss of resilience from 37 which is the area under the horizontal line (**Fig. 6**). If there was no disruptive event or if a county subdivision did not lose any resilience, population activity function would follow the horizontal line (baseline) in **Figs, 6-8**, resulting in no drop of activity with an activity ratio of 1. Since the considered study time is 37 days, the area under the population activity function (horizontal line) would be $(1 \times 37) = 37$. Thus, the maximum possible resilience (MPR) value should be 37. We also calculated the percentage loss of resilience, dividing the estimated TRL values by MPR value (MPR = 37). So, if a county-subdivision's TRL value is 12.88, the percentage loss of resilience will be 34.81% (12.88/37*100).

**4.2 Statistical modeling approach**

To determine the effects of different factors on the transient loss of community resilience, a Generalized Linear Mixed Model (GLMM) with Gamma family was developed. This study used GLMM for two reasons: (i) it is likely that the observations used in this study are not independent since the transient loss of resilience values for the subdivisions under a county might be similar to each other; in GLMM, to account for the non-independence issue, a random effect is introduced into the linear predictor of a regression model (**Equation 3** and **4**); (ii) **Fig. 11** shows that the transient loss of resilience is not normally distributed and it ranges between 0 and 37. So, community loss of resilience is skewed and always positive (TRL > 0). In GLMM, it is possible to account for any distribution (e.g., Gaussian, Beta, Poisson, Gamma etc.) of the dependent variable.

In general, the GLMM can be given by the following equations (*61*):

$$Mixed\ model = Fixed\ effect + Random\ effect \tag{3}$$

$$y = X\beta + Zb + \epsilon \tag{4}$$

where, $y$ is a $N \times 1$ column vector of continuous dependent variable, $X$ is a $N \times p$ matrix of the $p$ predictor variables; $\beta$ is a $p \times 1$ column vector of the fixed-effects regression coefficients; $Z$ is the $N \times q$ matrix for the $q$ random effects; $b$ is a $q \times 1$ vector of the random effects (the random complement to the fixed $\beta$); and $\epsilon$ is a $N \times 1$ column vector of the residuals. Here, the dependent variable $y$ can follow any distribution. Also, $b$ is not actually estimated; instead, $b$ is assumed as normally distributed with a zero mean and variance $G$ [i.e., $b \sim N(0, G)$]. Since the fixed effects are directly estimated, including the intercept, random effect complements are modeled as deviations from the fixed effect, so they have zero means. The random effects are just deviations around the value in $\beta$.

In this study, there were 166 observations ($N = 166$) from 36 parishes over 37 days. Since the dependent variable, transient loss of community resilience was continuous, always positive, and not normally distributed, the Gamma family of distribution was used with a log link function (*62*). Further, the model was specified with 8 fixed effects as predictor variables shown in **Table 1** and a random intercept for every county (parish). A parish had a random effect in the model since it was expected that due to spatial proximity the loss of resilience values of the county subdivisions in a parish would be correlated with each other. The model was estimated in R software and all the variables were standardized before fitting the model.

## 5. RESULTS

This section presents the results in two parts. First, it shows the visualization of our datasets and the quantified resilience and transient loss of resilience. Second, it presents the results of the Generalized Linear Mixed model (GLMM).

**5.1 Population Activity Trajectories**

Fig. 7 shows how population activity (in terms of activity on Facebook) fluctuated due to Hurricane Ida for 11 county subdivisions under 11 different parishes. It shows that the population activity rate started to decrease from 25th August 2021. The highest decrease in population activity was observed on 29th and 30th August 2021 (the day and the next day when the landfall occurred). The red vertical line in **Fig. 7** indicates the day when the landfall of Hurricane Ida occurred. By 30th September, all the county subdivisions (apart from county subdivision under Terrebonne) recovered. However, recovery times varied across county subdivisions. Some county subdivisions recovered fast (e.g., District 9 of East Baton Rouge parish). On the other hand, some county subdivisions had a longer time to return to a typical level of population activity (e.g., New Orleans). Similarly, maximum impact varied across county subdivisions. For example, in District 9 of Plaquemines parish, the population activity rate dropped below 0.2; on the other hand, in District 9 of East Baton Rouge, it dropped to 0.8 only.

**Fig. 8** shows the population activity rates at county subdivision level for Jefferson, Lafourche, Assumption and Plaquemines parishes. Population activity pattern under Hurricane Ida was homogeneous for Jefferson Parish. Population activity fluctuation curves in subdivisions under Jefferson parish followed very closely to each other. On the other hand, population activity fluctuation curves for subdivisions under Lafourche, Assumption and Plaquemines parishes were not exactly same but subdivisions under a parish followed a certain pattern most of the time. Similar to **Fig. 7**, the highest decrease in population activity was observed on 29th and 30th August 2021 in subdivisions under these parishes. Recovery time and maximum impact varied across subdivisions. Despite the differences in recovery time and maximum impact, the common property of the population activity fluctuation curves (**Fig. 7** and **8**) was: at the starting of a hurricane, population activities started to decrease in the affected area, had the maximum drop after a certain time (mostly on the landfall day and the next day), and then started to recover.

**5.2 Power Outage Trajectories**

**Fig. 9** shows the percentage of customers in different parishes who faced power outages due to Hurricane Ida. It shows that customers started to lose electricity supply after 28th August 2021. Most of the customers lost electricity services on 29th and 30th August 2021, the day, and the next day when landfall occurred. Similar to **Fig. 7**, the red vertical line in **Fig. 9** indicates the landfall day. In some counties (for example, Jefferson, St. Charles, Lafourche, Plaquemines, Terrebonne, St. John the Baptist, Orleans), about 100% customers lost power services. On the contrary, few customers from Iberia and Cameron lost their power services. It took long time for the customers in Terrebonne (28 days), Lafourche (29 days), and St. John the Baptist parishes (27 days) to restore the power services. Restoration time was shorter for the customers in St. Mary (3 days).

### 5.3 Community Resilience

**Tables 2** and **3** present the least resilient and most resilient county (parish) subdivisions in Louisiana due to Hurricane Ida giving the transient loss of resilience and resilience values, estimated from **Equation 2** (for more details see section 4.1). The transient loss of resilience indicates the area between population activity function Q (t) and baseline in **Figs. 6, 7** and **8**. Resilience is the area below the population activity function Q (t) of the same Figs. (section 4.1). The highest transient loss of resilience was found 12.88 for District 8 in Plaquemines parish. A higher value of transient loss of resilience indicates that people living in a county subdivision have lost much resilience in terms of their activity during Hurricane Ida. On the other hand, a higher value of resilience indicates that people living in a county subdivision were resilient to the hurricane. The TRL/MPR values are presented in the last columns of **Tables 2** and **3**.

During Hurricane Ida, among the 36 parishes in Louisiana considered here, the county subdivisions of Plaquemines suffered the highest transient loss of resilience followed by the county subdivisions of St. John the Baptist, Terrebonne, Lafourche, St. Charles, and Orleans (**Table 2**). These county subdivisions lost significant percentage of their resilience as indicated by TRL/MPR ranging from about 18% to 35% (**Table 2**). On the contrary, the county subdivisions under East Baton Rouge parish had the highest community resilience followed by the county subdivisions under Vermilion, Iberville, Iberia, and Lafayette parishes (**Table 3**). All transient loss of resilience values are very small (less than 1), and all the percentage loss of resilience values are less than 1.5% (**Table 3**) for these county subdivisions, indicating people living in these county subdivisions were resilient to the hurricane. Thus, these metrics could measure the extent of disruptions in population activity after Hurricane Ida.

### 5.4 Spatial Distribution of Community Resilience

**Fig. 10** shows the spatial distribution of transient loss of resilience over county subdivisions. It shows that county subdivisions in South-East Louisiana suffered higher transient loss of resilience. Transient losses of resilience were higher in the county subdivisions under Terrebonne, St. John the Baptist, Plaquemines, Lafourche, and St. Charles. At the time of landfall (29$^{th}$ August 2021), these places were close to the hurricane path. Since wind speed of hurricane path fell to 65 mph from 130 mph when it was over Livingston, some of the county subdivisions under this parish did not have higher transient loss of resilience, despite being close to hurricane path. Most of the county subdivisions from North (e.g., county subdivisions under St. Helena and Washington parish) and North-Western (e.g., county subdivisions from East Baton Rouge to far north Calcasieu parish) side of hurricane path resulted in lower loss of resilience (between 0 to 1). Although district 4 of Vernon parish and some of the county subdivisions from Cameron parish are far from hurricane path, they had moderate transient loss of resilience (between 4 and 6).

**Fig. 11** shows the distribution of the values of transient loss of resilience. Among 166 county subdivisions, about 60 county subdivisions had transient loss of resilience value between 0.1 to 1 and for 46 county subdivisions, the value was 4.

### 5.5 Result from the GLMM Model

**Table 4** presents the results of the GLMM model. Among multiple types of physical infrastructure disruption related predictor variables, disruption to transportation systems and power outage were found to be significant and positively associated with transient loss of resilience. A positive association means that an increase in a predictor variable will increase the loss in resilience and a negative association indicates the opposite. The exponentiated coefficient of duration of disruptions on roads (e^{0.141} = 1.1514) is the factor by which the mean transient loss of resilience increases by 1.15 times with one hour increase in duration of disruptions to transportation network. One day increase in restoration time for power outage (e^{0.632} = 1.88) increases the mean transient loss of resilience by 1.88 times. Housing damages had positive association with transient loss of community resilience too, but this predictor variable was not found to be significant.

Distance to hurricane path was found to be significant and positively associated with loss of community resilience. In addition, median household income was found to be significant among the socio-economic characteristics of the communities. The percentage of Black and Hispanic population and the percentage of houses built before 2000 had negative association with loss of resilience, but those were not found to be statistically significant.

The variance for parishes (random effects) was found to be 0.1185 and 0.325 for residuals. The random effects are important as they explain a significant amount of variation. We can take the variance for the parishes and divide it by the total variance: [0.1185/ (0.1185 + 0.3250)] = 27%. So, the differences between parishes explain ~27% of the variance that has been left after explained by the fixed effects. The $R^2$ (marginal), representing the proportion of variance explained by the fixed effects, has a value of 0.44. The $R^2$ (conditional), representing as the proportion of variance explained by the entire model, including both fixed and random effects, has a value of 0.61. Therefore, due to introducing parishes as random effects, the proportion of variance explained by the model increased.

### 6. DISCUSSIONS

This study used large-scale Facebook population data for Hurricane Ida occurred at Louisiana, to explore how population activity before, during and after a disaster can quantify the resilience and transient loss of resilience of the affected communities. Since Facebook population dataset provides data at county subdivision level, this work quantifies the loss of community resilience at a higher resolution, whereas previous studies mainly focused on county level macroscopic analysis. It was found that subdivisions under Plaquemines, Lafourche, St. John the Baptist, Orleans, Terrebonne, St. Charles, and Jefferson parishes were more negatively impacted by Hurricane Ida.

This paper also studied how the transient loss of community resilience is associated with the disruptions in multiple physical infrastructure systems, disaster condition, and socio-economic characteristics from a Generalized Linear Mixed Model (GLMM). A positive coefficient for duration of disruption on roads indicates that the transient loss of resilience was higher in places having a longer duration of road blockage. This implies that people who evacuated could not immediately return to their homes; and as a result, their regular activities were not observed shortly after the hurricane. The importance of transportation networks on the recovery of disaster affected regions was understudied, despite having significant implications on policymaking for community resilience. This study suggests that performance of transportation systems should be enhanced before and after hurricanes to ensure fast evacuation and recovery of a region. Similar to duration of disruption on roads, the positive coefficient for the restoration time of power outage indicates that a longer restoration time in power services causes a higher loss of resilience. This implies that people might have connectivity issue in their area that prevented their normal social activity which resulted in a higher loss of resilience. Therefore, to enhance community resilience against an extreme event, faster restoration from power outages should be a necessary recovery effort. Power service restoration time is associated with recovery speed of population activity as found in disasters including Hurricanes Irma and Maria (*16*).

This study found a positive coefficient between the distance to hurricane path and the transient loss of resilience. This may appear counterintuitive as it indicates that regions farther from the hurricane path would suffer a higher loss of resilience. However, this is probably because of the presence of the variable indicating the power outage restoration time in the model. The positive coefficient for the distance to hurricane path implies that if two regions faced same restoration time of power outage but located at different distances from hurricane path, the region which is located far away from the hurricane path, suffers a higher transient loss of resilience (*63*). A possible reason for this could be that places which were close to hurricane path were given priority including financial and logistical support during recovery process from different humanitarian and relief organizations or those communities might have better disaster preparedness because those places are prone to hurricanes. To further investigate this issue, we explored data from Community Emergency Response Team (CERT) Dataset (*64*). This Program educates people about disaster preparedness for hazards that may impact their area and trains them in basic disaster response skills. Among the considered 36 parishes for Hurricane Ida, CERT has programs in only 12 parishes in Louisiana and majority of those parishes (10 out of 12) are located close to the Hurricane Ida's path. This implies that disaster preparedness programs can be expanded among communities who live far away from the hurricane path to educate and train them about the basic disaster response skills about team organization, and medical operations. This also indicates that "disaster preparedness training" component of **Fig. 2** contributes to strengthen the community resilience.

Median household income had negative effects on transient loss of community resilience, implying that communities with a lower median income had a higher loss in resilience. In other words, communities with lower economic resources faced greater challenges in recovering and adapting to the impact of the hurricane. This could be since low-income communities experienced difficulties in their regular activity for a longer time as regions with poorer communities might have less robust infrastructure systems for post-disaster recovery (*16*). Low-income communities might have weaker governmental support and fewer financial and material resources to respond effectively to a disaster. This could hinder their ability to access emergency services, repair infrastructure, and provide necessary support to residents. Low-income communities might also face greater social vulnerabilities, such as lack of access to healthcare, limited transportation options, and inadequate housing due to property damage. These factors can amplify the negative effects of a disaster and prevent fast recovery (*16, 40, 65, 66*).

This study found that percentage of Hispanic and Black population in different county subdivisions had a negative impact on loss of community resilience. However, these variables were not found to be statistically significant, indicating a lack of evidence to reject the hypothesis that ethnicity is not associated with loss of community resilience during Hurricane Ida in Louisiana. However, previous studies found evidence of inequality in community resilience and experienced hardship in Texas during Hurricane Harvey (*8, 23*) as poor and minority communities were less prioritized in recovery efforts. Areas with less vulnerable people (in terms of ethnicity, income, age, gender, employment status) recovered faster than areas with more vulnerable populations in New Orleans (*39*). Also, applicants and beneficiaries in South Carolina for the 2015 floods and 2016 Hurricane Matthew were among the most socially vulnerable within a census tract (*67*). Cutter et al. (*68*) found that socially vulnerable population, despite not residing in the highest areas of disaster risks, can undergo long-term recovery from disasters. Besides recovery after landfall, migration (*69*) and evacuation tendency (*70*) were found to be negatively associated with median income during Hurricane Katrina and Rita.

Similarly, this study also found disparity issues in Louisiana due to Hurricane Ida. The negative relationship between median household income and loss of community resilience indicates an inequality issue. Accelerated recovery efforts and better infrastructure systems are needed in low-income communities to make them resilient to hurricanes, highlighting the need for targeted policies and interventions that can address their specific challenges during disaster recovery. It underscores the importance of considering socioeconomic factors when planning for disaster preparedness and response as vulnerable communities are more likely to experience prolonged recovery periods. Disaster management agencies should ensure that resources are allocated equitably to communities with varying income levels to minimize disparities in resilience outcomes.

The described approaches and findings of the study can benefit policy making and disaster management in several ways. First, the proposed method to quantify '*transient loss of resilience*' using Facebook data can be used to better understand the negative impacts of a disaster to an affected community since it involves both systemic impact (e.g., percentage of population impacted by a disaster) and time to recover. Second, since Facebook data is globally available for the affected region, it can be widely used by policy makers and disaster management agencies to understand the negative impact on the affected community by quantifying the loss of community resilience. Third, since this study analyzed the data at a higher resolution (county subdivision level) compared to a county level analysis, the findings ensure better understanding of community resilience as well as the association between loss of community resilience and physical infrastructure disruptions, disaster condition and economic aspects. Fourth, the findings of this study suggest that the association between social and physical systems should be considered to strengthen community resilience. Due to the association between community resilience in terms of population activity and physical infrastructure systems, infrastructure disruptions caused by disasters can exacerbate the hardship experienced by the affected population. Lastly, the findings also suggest that better and fast recovery policies, and better infrastructure systems should be given emphasis by the policy makers in less wealthier communities to make them resilient to hurricanes.

Since this study proposes methodology to use Facebook data, we mention the strengths and limitations of using this data as follows:

**Strengths of Facebook data**

Previously used high-resolution location data including taxi data (*9*, *10*), GPS data (*11*, *12*), Wi-Fi (*14*), and mobile phone datasets (*13*, *15*, *16*) are often proprietary or accessible to only few researchers. To effectively quantify community resilience, data must be easily accessible and usable (*17*). These data unavailability issues can be avoided if location data from a widely used service is made accessible. Facebook population data provides aggregate information of Facebook users following a crisis event to help humanitarian organizations (*25*). Data from this platform are free and easily available to researchers and policymakers; whenever there is a disaster, such aggregate data is globally available from the affected regions. As such, this dataset is a great source for decision makers to identify crisis events over a certain time window and investigate the impacts of these events on population activity.

Previous research on social media data analyzed mainly social media posts (text data) or check-in data. However, few people post or check-in on social media (Facebook and Twitter) during a disaster; and among the posted tweets and Facebook status on average 1-2% data has specific location information (*71*). Thus, social media posts or check-in data may result in a small penetration rate of actual population. Facebook data does not require users to post or check-in, ensuring a higher percentage of sample size/penetration to actual population. Another limitation of using social media data is that different studies estimated location data metrics such as baseline population in different ways; no universal assumption exists to extract such metrics from the raw data so that disaster officials can widely use social media data for crisis management. Facebook data gives such a metric by setting fixed assumptions to be used by all data users; this also minimizes required data processing steps to be performed by disaster officials. As such, this dataset is available with a reasonable sample size and useful location information in a usable format that requires less data processing efforts and assumptions.

**Limitations of Facebook data**

Facebook data may have limits due to self-selection. Older population may have less interest in using social media or younger persons may use other social media platforms such as TikTok, Twitter, or Instagram more than Facebook. In such cases, Facebook data may have smaller sample size for some population groups. Ribeiro et al. (*72*) found a higher percentage of Facebook users among different races in data from Facebook Marketing API platform. However, Facebook population data from 'Facebook Data for Good' platform gives aggregate statistics without any personal information of Facebook users to protect user privacy. As such, it is not possible to estimate the penetration rate to population of a particular demographic (e.g., age, gender, race, income) in an affected area. Facebook data can have overall larger sample size but smaller penetration from a particular group of people, potentially introducing biases in analysis results.

Overall, emergency officials and decision-makers can rely on Facebook population data from 'Facebook Data for Good' platform due to its open accessibility, larger sample size, higher correlation between sample size and actual population, and universal assumption to interpret location data metrics.

**7. CONCLUSIONS**

This study quantified transient loss in community resilience and used it effectively to identify less resilient (more negatively impacted) communities to hurricanes from large-scale, real-time, free, and easily accessible Facebook data. The use of large-scale location data enables proactive monitoring of population activity before, during, and after a disaster such that the impact to affected community can be evaluated in real-time. This study can be used as a reference for local governments and policymakers to decide equitable spatio-temporal allocation of resources and services like food, utilities, and optimized shelter locations by rapid impact assessment based on observed loss of resilience shortly after a disaster. This study also investigated why some communities are more impacted. Transient loss in community resilience was examined for Hurricane Ida using generalized linear mixed models at county subdivision level. Using models, descriptive statistics, and geospatial analytics, this study identified consistent relationships between transient loss of community resilience and key factors: multiple types of physical infrastructure disruptions, disaster condition, and socioeconomic characteristics. This study found a disparity issue in recovering after a hurricane suggesting that communities with lower income were more impacted. This emphasizes accelerated recovery efforts and better infrastructure systems in low income communities.

Loss in community resilience indicates drop in regular population activity. A decline in population activity has a ripple effect on many aspects of society, from banking and finance to education and healthcare. Thus, policy makers and disaster management officials should effectively identify impacted communities in real-time and accelerate the recovery process in transportation and power infrastructures, provide equal recovery services and pre-disaster preparedness trainings across all communities without considering the economic characteristics. As a result, this will help make communities more resilient to hurricanes.

The study has some limitations: (i) this study assumed that all county subdivisions under a parish had the same power outage restoration time as electricity companies do not reveal the postal code and city name in the provided datasets. Depending on data availability, future studies can consider power outage data at county subdivision level. (ii) If a specific community doesn't access to Facebook (such as older people, people in rural areas, or use other social media rather than Facebook), this dataset may result in small sample size (small penetration rate to actual population) and may not have uniformly distributed users from all demographics. Data should be carefully used in such cases and maybe some other data platform can be used in combination with Facebook Data for Good platform.

Future studies can focus on the biases in social media based location data. Using Facebook data future studies can also focus on if community resilience can be identified by community usage of infrastructure services or how different is that from population activity revealed by Facebook data. Due to the interaction between infrastructure systems and population activity on social media, these anticipated post-disaster activity curves can be used by emergency personnel and policymakers as an indicator of the spatio-temporal pattern of electricity and other infrastructure disruptions.

**DATA AVAILABILITY STATEMENTS**

All data, models, or code that support the findings of this study are available from the corresponding author upon reasonable request. All data sources have been mentioned in Data Description chapter.

**ACKNOWLEDGMENTS**

The authors are grateful to the U.S. National Science Foundation for the grants EAGER SAI–2122135 and CMMI-1832578 to support the research presented in this paper. However, the authors are solely responsible for the findings presented here.

**FIGURE CAPTIONS**

**Fig. 1.** Influence of decreased human activity on daily life.

**Fig. 2.** Factors associated with decreased population activity.

**Fig. 3.** Facebook users vs population in 327 parish subdivisions and 39 parishes of Louisiana.

**Fig. 4.** Power outage for Livingston Parish due to Hurricane Ida and considered restoration time in this study.

**Fig. 5.** Correlations among variables

**Fig. 6.** Conceptual definition of resilience and transient loss of resilience for the affected communities.

**Fig. 7.** Population activity curves from Facebook population datasets in Louisiana due to Hurricane Ida.

**Fig. 8.** Population activity curves for subdivisions under four different parishes

**Fig. 9.** Power outage curves due to Hurricane Ida.

**Fig. 10.** Transient loss of resilience for 166 county subdivisions along with hurricane path.

**Fig. 11.** Distribution plot of transient loss of community resilience.

# TABLES

## Table 1. Descriptive Statistics

| Variables | Mean | Std | Min | Median | Max |
|---|---|---|---|---|---|
| Transient loss of resilience (activity ratio-day) | 3.19 | 2.81 | 0.101 | 2.278 | 12.882 |
| **Physical infrastructure data** | | | | | |
| Duration of disruption on roads (hr) | 490.07 | 1202.94 | 0.5 | 76.55 | 11441.32 |
| Restoration time for power outage (Day) | 12.49 | 9.13 | 0 | 13 | 29 |
| Property damage | 4051.81 | 7045.78 | 0 | 1539 | 45117.71 |
| % of Households built before 2000 | 72.5 | 17.95 | 0 | 75.45 | 96.4 |
| **Disaster condition** | | | | | |
| Distance to hurricane path (km) | 67.6 | 55 | 5.734 | 49.067 | 287.38 |
| **Socio-economic characteristics and social vulnerability** | | | | | |
| Median household income (USD) | 53791.48 | 19837.67 | 16583 | 52085 | 133056 |
| % of Black population | 29.7 | 23.73 | 0 | 23.8 | 93.4 |
| % of Hispanic population | 3.91 | 3.73 | 0 | 2.90 | 20.7 |

**Table 2.** Transient loss of resilience and resilience values of 20 least resilient county subdivisions

| Parish Name | County (Parish) subdivision | Transient loss of resilience (TRL) | Remaining Resilience | (TRL/ MPR) *100 % |
|---|---|---|---|---|
| **Plaquemines** | District 8 | 12.88 | 24.12 | 34.81 |
| | District 9 | 12.22 | 24.72 | 33.03 |
| | District 7 | 11.62 | 25.39 | 31.41 |
| | District 6 | 9.61 | 27.39 | 25.97 |
| **St. John the Baptist** | District 5 | 11.07 | 25.93 | 29.92 |
| | District 7 | 8.63 | 28.37 | 23.32 |
| **Terrebonne** | District 7 | 10.94 | 26.06 | 29.57 |
| | District 1 | 9.72 | 27.28 | 26.27 |
| | District 8 | 8.57 | 28.44 | 23.16 |
| | District 9 | 8.52 | 28.48 | 23.03 |
| **Lafourche** | District 9 | 10.07 | 26.93 | 27.22 |
| | District 5 | 7.84 | 29.16 | 21.19 |
| **St. Charles** | District 3 | 8.72 | 28.28 | 23.57 |
| | District 2 | 7.72 | 29.28 | 20.86 |
| | District 6 | 7.70 | 29.30 | 20.81 |
| **Sabine** | District 1 | 8.28 | 28.72 | 22.38 |
| **Orleans** | New Orleans | 8.28 | 28.72 | 22.38 |
| **Livingston** | District 8 | 6.87 | 30.13 | 18.57 |
| **Cameron** | District 1 | 6.80 | 30.20 | 18.38 |
| **Jefferson** | District 5 | 6.76 | 30.24 | 18.27 |

MPR = Maximum Possible Resilience, which is 37 as described in section 4.1

**Table 3.** Transient loss of resilience and resilience values of 20 resilient county subdivisions

| Parish Name | County (Parish) subdivision | Transient loss of resilience (TRL) | Remaining Resilience | (TRL/ MPR) *100 % |
|---|---|---|---|---|
| **East Baton Rouge** | District 4 | 0.10 | 36.9 | 0.27 |
| | District 3 | 0.15 | 36.85 | 0.41 |
| | District 12 | 0.30 | 36.7 | 0.81 |
| | District 7 | 0.49 | 36.51 | 1.32 |
| **Vermilion** | District 13 | 0.26 | 36.74 | 0.70 |
| **Ascension** | District 8 | 0.29 | 36.71 | 0.78 |
| | District 10 | 0.36 | 36.64 | 0.97 |
| **Iberville** | District 13 | 0.30 | 36.70 | 0.81 |
| | District 10 | 0.30 | 36.70 | 0.81 |
| **Iberia** | District 5 | 0.34 | 36.66 | 0.92 |
| | District 13 | 0.52 | 36.48 | 1.41 |
| **St. Tammany** | District 8 | 0.36 | 36.64 | 0.97 |
| | District 9 | 0.50 | 36.50 | 1.35 |
| **Washington** | District 3 | 0.39 | 36.41 | 1.05 |
| **Livingston** | District 3 | 0.34 | 36.66 | 0.92 |
| | District 5 | 0.40 | 36.60 | 1.08 |
| | District 7 | 0.46 | 36.54 | 1.24 |
| **Lafayette** | District F | 0.43 | 36.57 | 1.16 |
| **St. Martin** | District 6 | 0.45 | 36.55 | 1.22 |
| | District 2 | 0.50 | 36.50 | 1.35 |

MPR = Maximum Possible Resilience, which is 37 as described in section 4.1

**Table 4.** Results of the generalized linear mixed model

| Variables | Estimate | Std. Error | t-statistics | p-value |
|---|---|---|---|---|
| Intercept | 0.936 | 0.107 | 8.697 | < 2e-16*** |
| **Physical infrastructure** | | | | |
| Duration of disruption on roads | 0.141 | 0.066 | 2.140 | 0.032** |
| Restoration time for power outage | 0.632 | 0.139 | 4.539 | 5.65e-06 *** |
| Property damage | 0.090 | 0.066 | 1.355 | 0.176 |
| % of houses built before 2000 | -0.092 | 0.070 | -1.316 | 0.188 |
| **Disaster condition** | | | | |
| Distance to hurricane path | 0.322 | 0.110 | 2.920 | 0.004 ** |
| **Socio-economic characteristics** | | | | |
| Median household income | -0.111 | 0.061 | -1.842 | 0.065 * |
| % of Black population | -0.006 | 0.070 | -0.086 | 0.932 |
| % of Hispanic population | -0.029 | 0.0515 | -0.555 | 0.579 |
| $R^2$ (marginal) | 0.44 | | | |
| $R^2$ (conditional) | 0.61 | | | |
| AIC | 612 | | | |
| BIC | 646 | | | |
| Log-likelihood | -295 | | | |
| Variance (parishes) | 0.1185 | | | |
| Residual variance | 0.3250 | | | |
| Significance level: <0.0001-'***', 0.001-'**', 0.05-'*', 0.1-'', 1 | | | | |

**FIGURES**

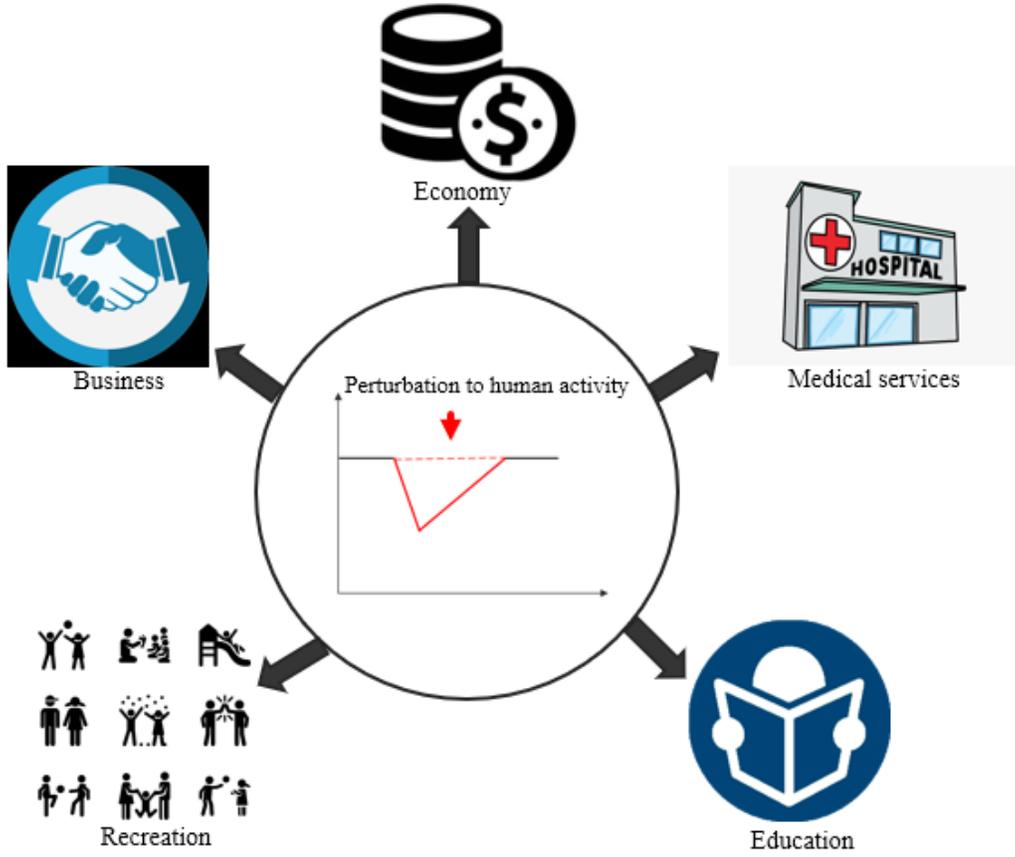

**Fig. 1.** Influence of decreased human activity on daily life.

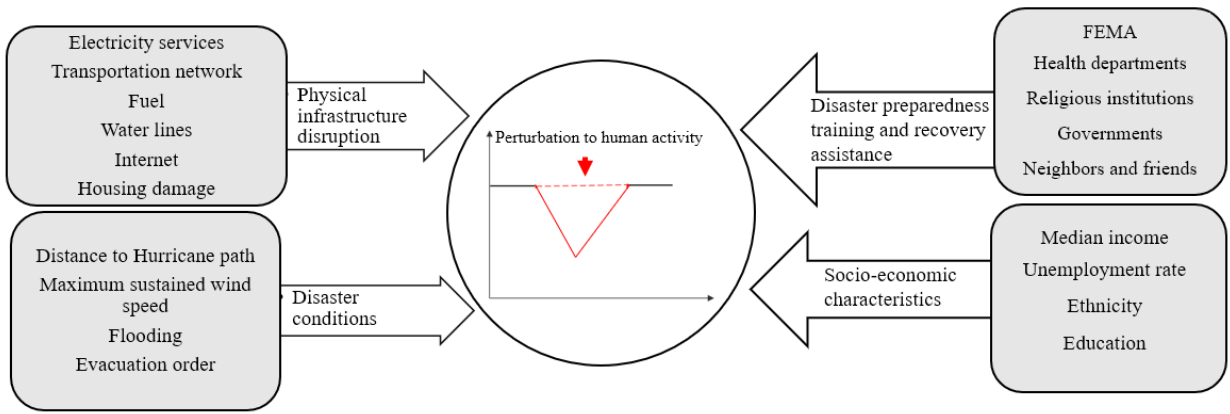

**Fig. 2.** Factors associated with decreased population activity.

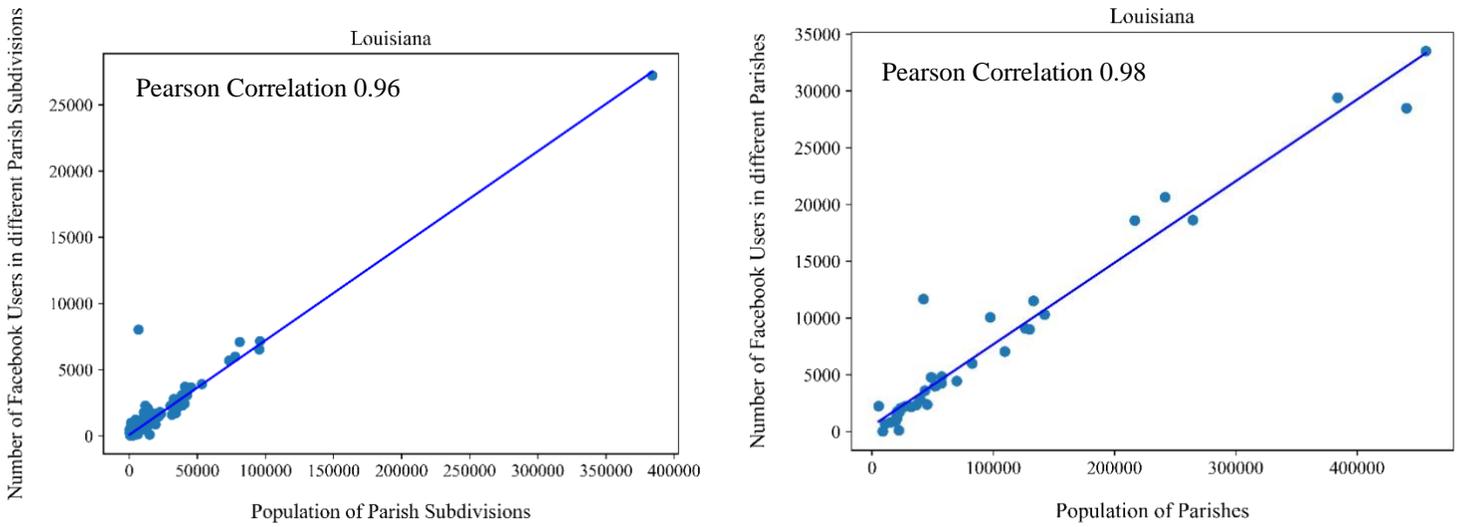

**Fig. 3.** Facebook users vs. population in 327 parish subdivisions and 39 parishes of Louisiana.

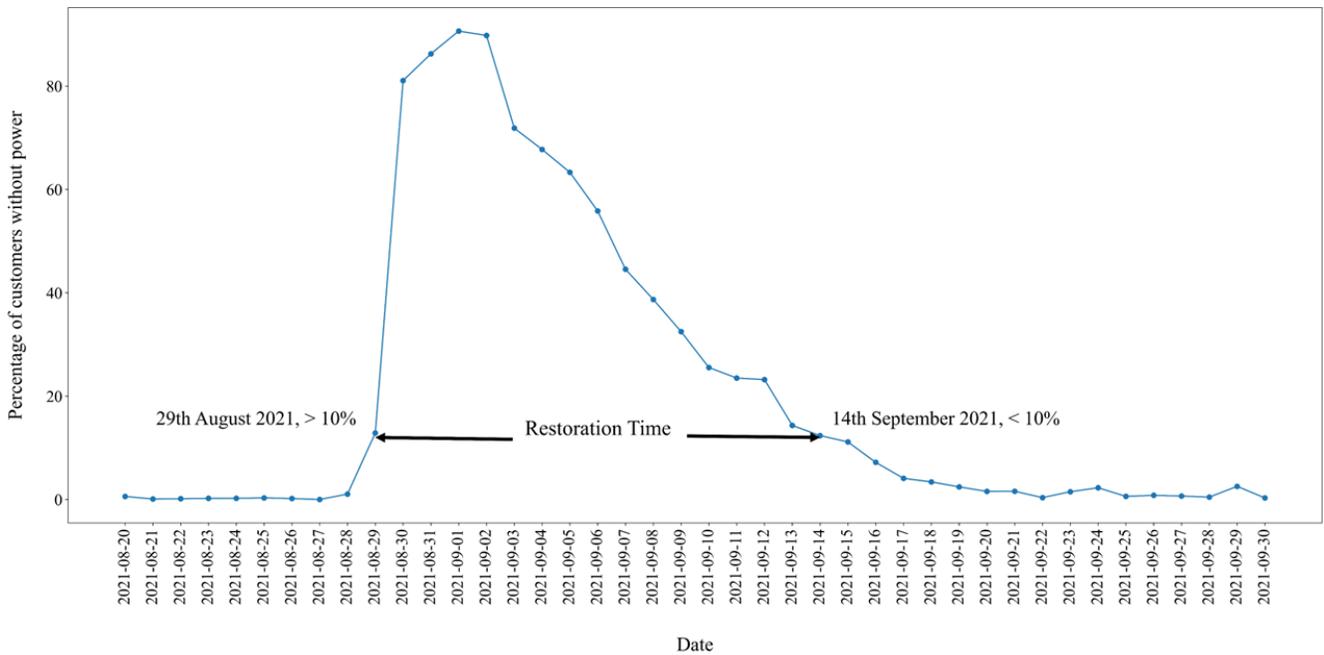

**Fig. 4.** Power outage for Livingston Parish due to Hurricane Ida and considered restoration time in this study.

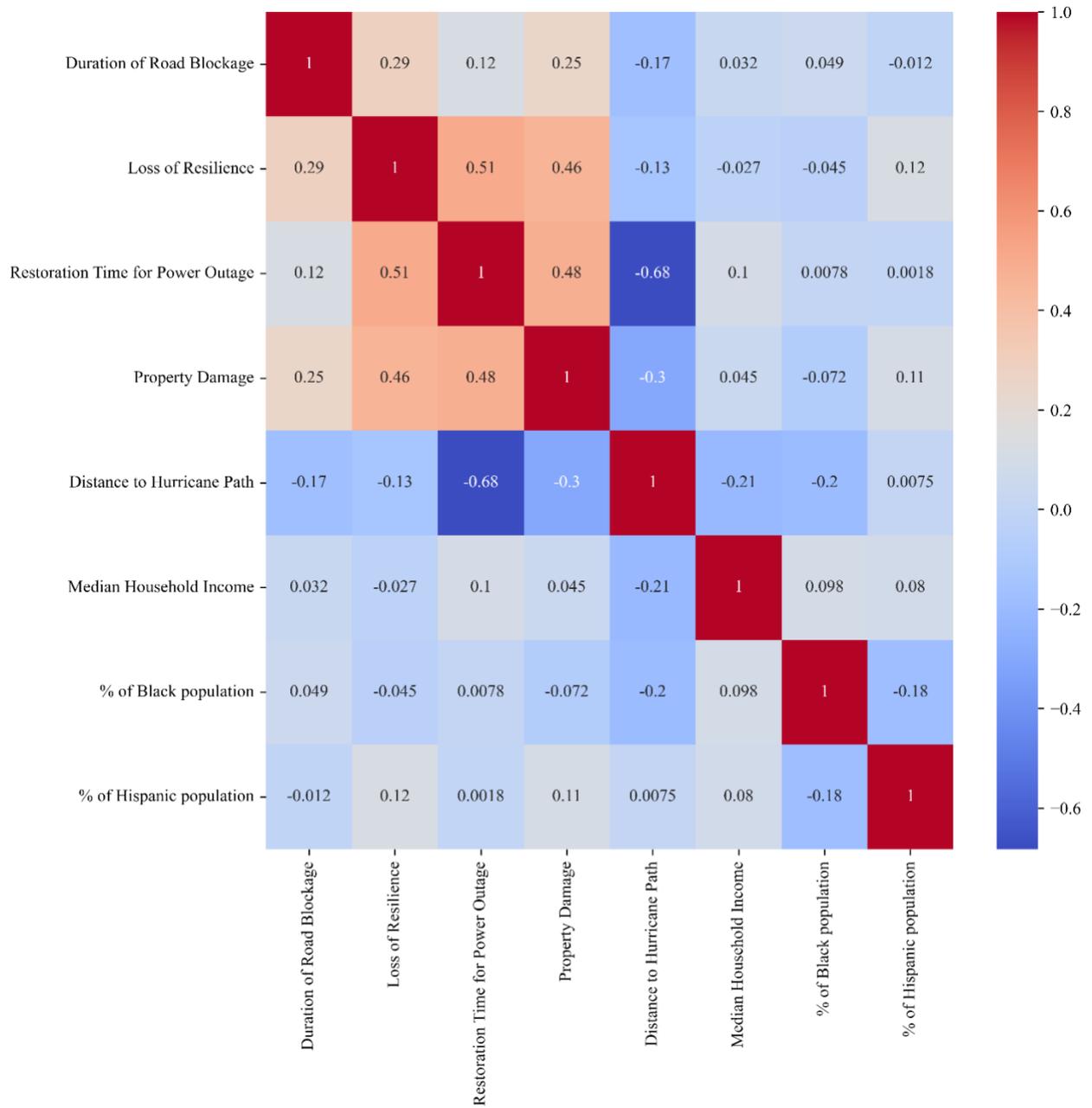

**Fig. 5.** Correlations among variables

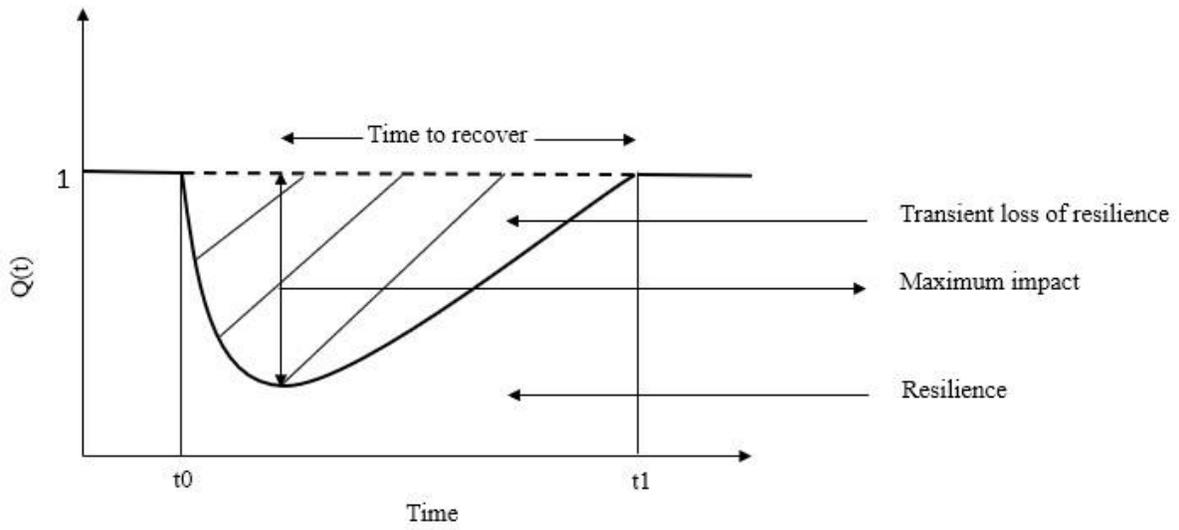

**Fig. 6.** Conceptual definition of resilience and transient loss of resilience for the affected communities.

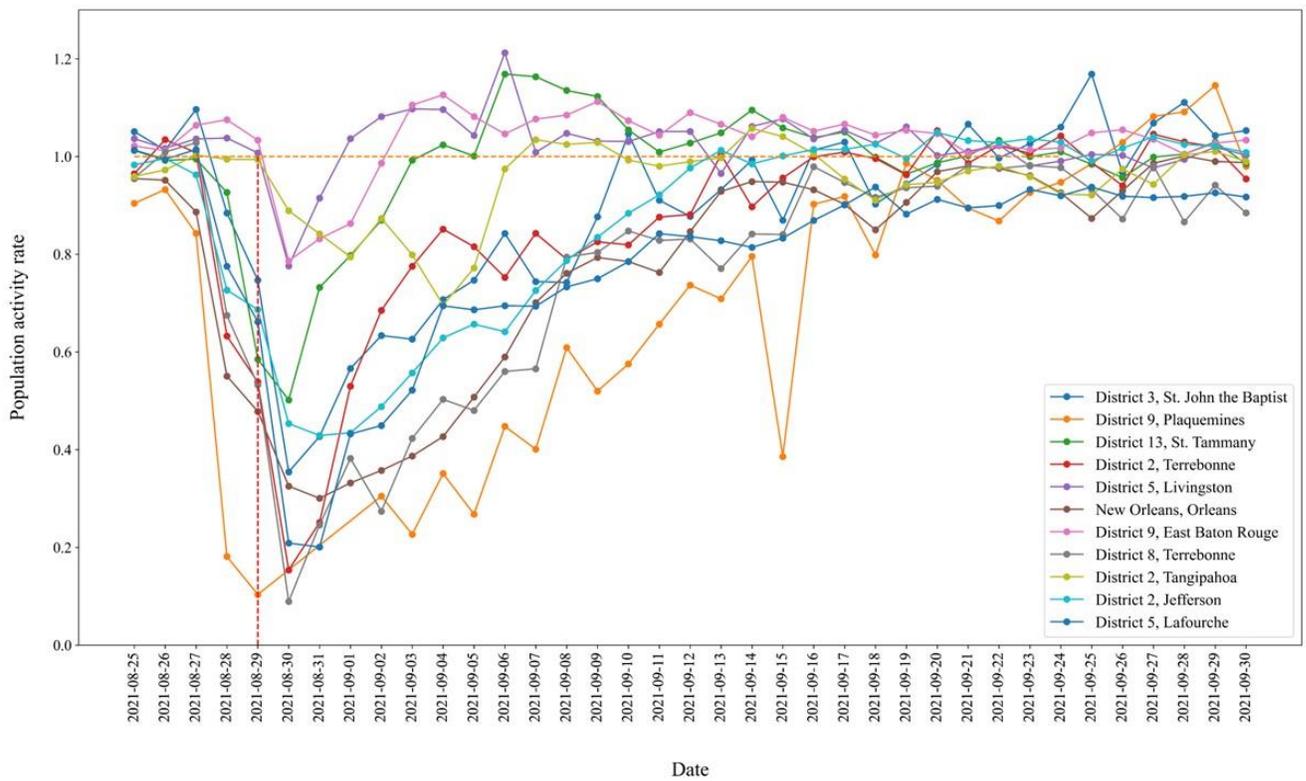

**Fig. 7.** Population activity curves from Facebook population datasets in Louisiana due to Hurricane Ida.

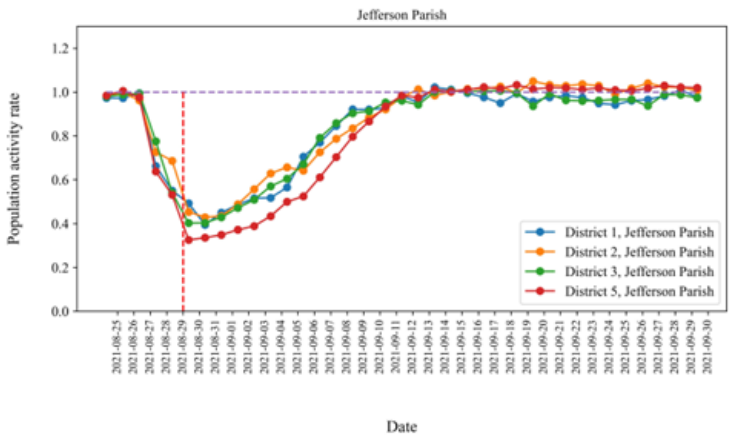
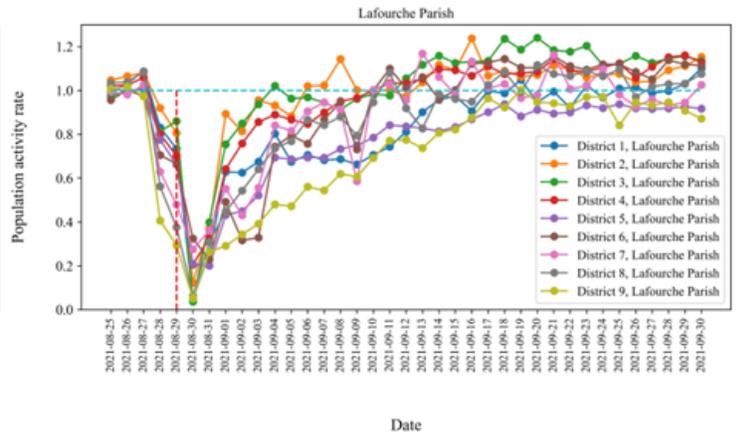
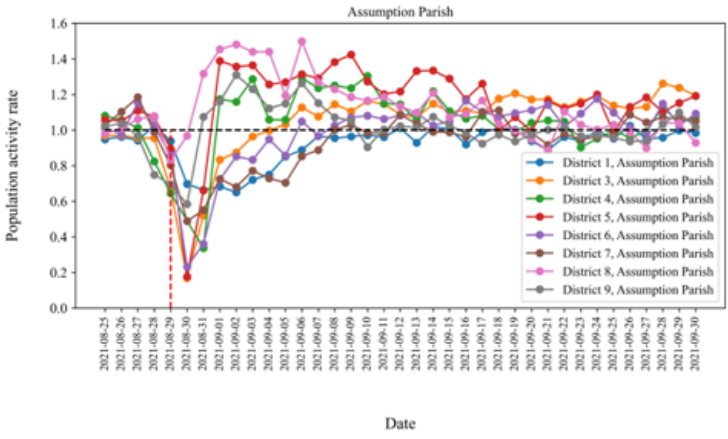
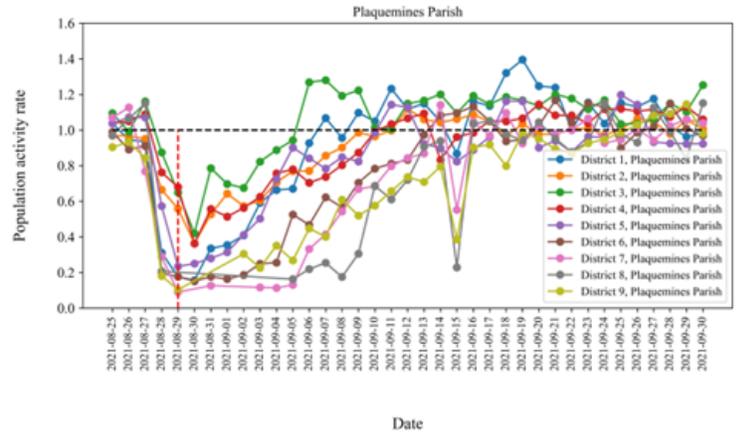

**Fig. 8.** Population activity curves for subdivisions under four parishes.

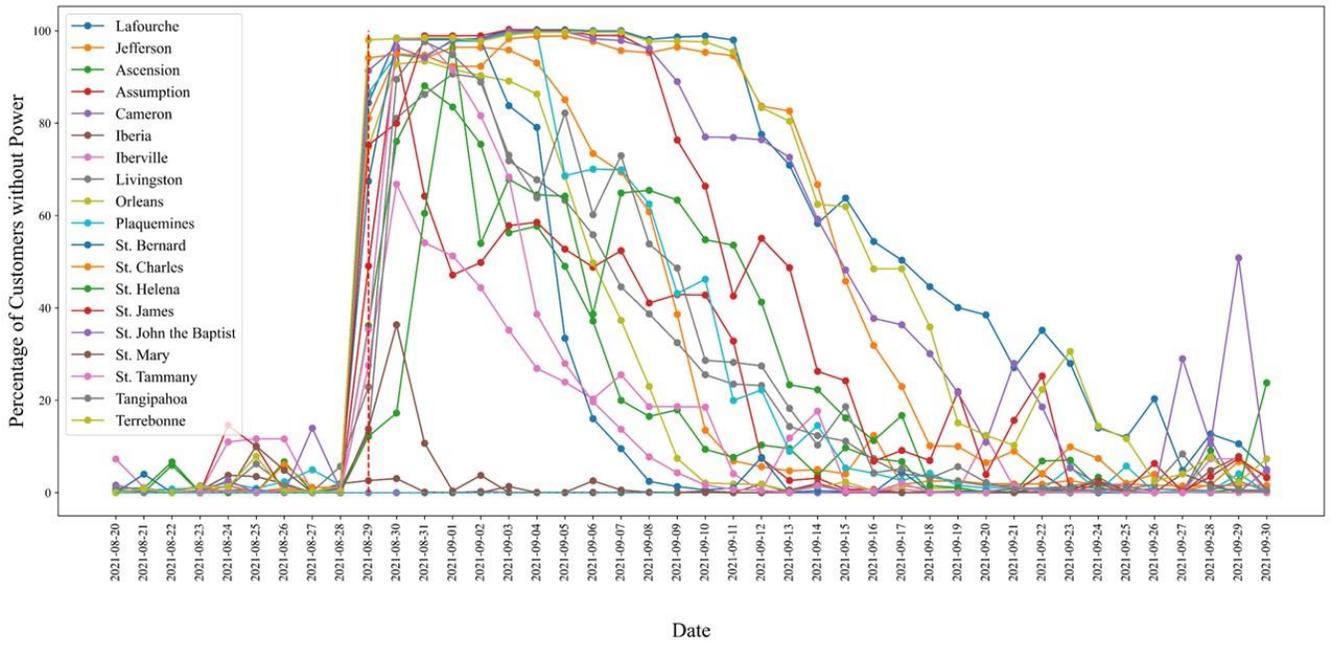

**Fig. 9.** Power outage curves due to Hurricane Ida.

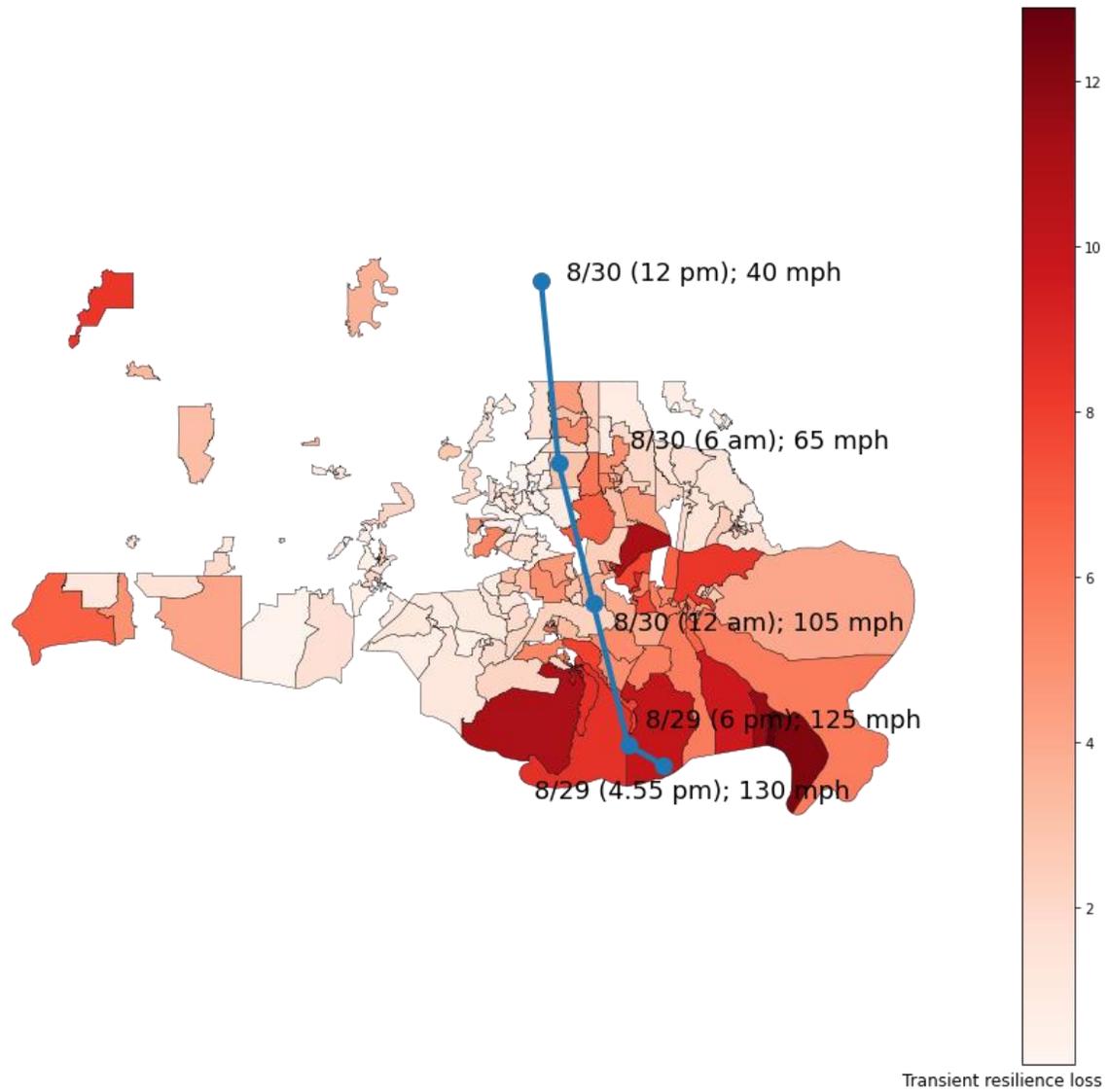

**Fig. 10.** Transient loss of resilience for 166 county subdivisions along with hurricane path.

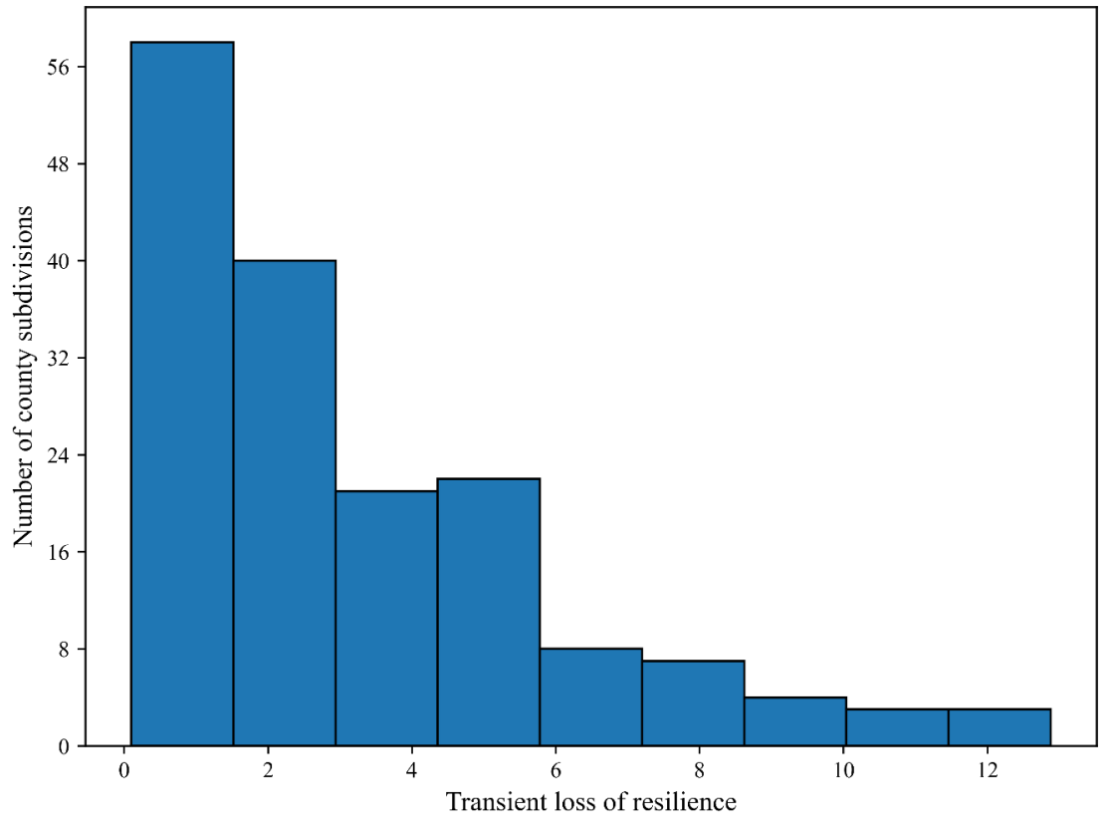

**Fig. 11.** Distribution plot of transient loss of community resilience.